\documentclass[twocolumn,aps,amsmath,amssymb,prb]{revtex4}
\usepackage{graphicx}
\usepackage{epsfig}
\bibstyle{apsrev.bib}

\begin{document}
\input epsf.sty

\title{Partially Asymmetric Exclusion Processes with Sitewise Disorder}

\author{R\'obert Juh\'asz}
 \email{juhasz@lusi.uni-sb.de} 
\affiliation{Fachrichtung Theoretische Physik, Universit\"at des
  Saarlandes, D-66041 Saarbr\"ucken, Germany}
\author{Ludger Santen}
 \email{santen@lusi.uni-sb.de}
\affiliation{Fachrichtung Theoretische Physik, Universit\"at des
  Saarlandes, D-66041 Saarbr\"ucken, Germany}
\author{Ferenc Igl\'oi}  \email{igloi@szfki.hu}
\affiliation{ Research Institute for Solid
State Physics and Optics, H-1525 Budapest, P.O.Box 49, Hungary}
\affiliation{ Institute of Theoretical Physics, Szeged University,
H-6720 Szeged, Hungary}

\date{\today}

\begin{abstract}
We study the stationary properties as well as the non-stationary dynamics
of the one-dimensional partially asymmetric exclusion process with
position dependent random hop rates. In a finite system of $L$ sites the
stationary current, $J$, is determined by the largest barrier and the corresponding
waiting time, $\tau \sim J^{-1}$, is related to the waiting time of a single random walker, $\tau_{rw}$,
as $\tau \sim \tau_{rw}^{1/2}$. The current is found to vanish as:
$J \sim L^{-z/2}$, where $z$ is the dynamical exponent of the biased single particle Sinai walk.
Typical stationary states are phase separated: At the largest barrier
almost all particles queue at one side and almost all holes are at the other side. 
The high-density (low-density) region,
is divided into $\sim L^{1/2}$ connected parts of particles (holes) which are separated
by islands of holes (particles) located at the subleading barriers (valleys).
We also study non-stationary processes of the system, like coarsening and invasion.
Finally we discuss some related models, where particles of larger size 
or multiple occupation of lattice sites is considered. 
\end{abstract}

\maketitle

\newcommand{\bc}{\begin{center}}
\newcommand{\ec}{\end{center}}
\newcommand{\be}{\begin{equation}}
\newcommand{\ee}{\end{equation}}
\newcommand{\beqn}{\begin{eqnarray}}
\newcommand{\eeqn}{\end{eqnarray}}

\vskip 2cm

\section{Introduction}

The stochastic dynamics of self-driven particles on one-dimensional lattices is
one of the key problems of non-equilibrium physics\cite{schmittman,schutzreview}. Models of this kind
have been used in order to describe such different problems as e.g. highway 
traffic\cite{chowd} or the dynamics of motor proteins along actin filaments or 
microtubuli\cite{mol_motors}. These two examples have in common that the directed 
particle motion against dissipative forces is maintained by the steady 
input of energy, a feature that leads to the generic non-equilibrium 
behaviour of the system. The finite drift of the particles leads to 
a strong  sensitivity to spatial inhomogeneities.
In contrast to equilibrium systems, even a single local defect may lead to 
bulk effects as e.g. the separation into macroscopic low and high density 
domains\cite{janowsky}. Moreover disorder of any type may also strongly influence the 
transport capacities of the system\cite{krug,barma_rw}. This is in particular true for the 
case of strong disorder, i.e. realizations of the disorder where the local
direction of the bias is non-uniform\cite{evans,barma,jsi,zrp}. The prototype of this kind of stochastic 
motion is the Sinai walk, which has been studied to great extend\cite{sinai}.
The stochastic motion of a Sinai walker is characterized by large velocity 
fluctuations. The origin of these fluctuations is most easily understood if 
one translates the local hop rates into energy differences, i.e. a forward bias 
corresponds to a negative slope of the corresponding energy landscape and
vice versa. Obviously the energy landscape is not flat for 
strong disorder, but characterized by barriers of different heights. 
The time a walker spends in front of such a barrier increases
exponentially with the height of the barrier and therefore the 
largest barriers determine the behavior of the system. 

There exists a large class of both quantum and stochastic models 
where the disorder plays a dominant role over deterministic fluctuations and where 
a connection to the Sinai walk can be made\cite{im}. In this respect one can
mention random quantum spin chains\cite{DF,fisherxx} or reaction diffusion models
with quenched disorder\cite{hiv}. Besides  a number of 
experimental setups have been identified, where this model 
properly describes the particle dynamics, e.g. the translocation of a 
RNA strand through a pore\cite{translocation} or the motion of molecular motors on 
microtubules\cite{mol_motors}.

While most of the models refer to the single particle case,
more recently, the effect of strong disorder on driven-diffusive 
many particle systems, i.e. the so-called asymmetric exclusion process
with strong disorder, has been
investigated\cite{evans,barma,derrida,stinchcombe,evans1,jsi,zrp}. It has been shown that 
the properties of the system largely depend on the way how the disorder 
is implemented. In case of particle disorder, i.e. for the case where 
the hopping rates of the particles are quenched random variables\cite{evans}, the
transport properties of the system are in close analogy 
to the single particle system\cite{jsi,zrp} both in the Griffiths phase, i.e. in case 
of a tilted energy landscape, and at the critical point, i.e. 
for a vanishing drift velocity of the particles. This 
implies that the many particle effects are most pronounced for quantities that 
are related to the arrangement of the particles. In case of lattice disorder,
however, the transport capacity of single and many particle systems
are also different\cite{barma,jsi}, because the presence of many particles alters the 
effective heights of the energy barriers. A relation between the stationary current
for particle disorder and for lattice disorder is announced in Ref.[\onlinecite{jsi}].
In the present work we elaborate the properties of the distribution of the stationary current
and show connections with extreme value statistics\cite{galambos,jli06}. We also
discuss the properties of the density profile as well as the consequences for the coarsening
behavior. Some of the results obtained in this work may also be relevant 
 for models which are closely related to the exclusion process, such as
the Heisenberg chain\cite{gwa_spohn} and the dimer evaporation and
deposition process \cite{stinchcombe1}.  

The article is organized as follows. The model is introduced in Sec.\ref{model} and
the basic properties of the single particle motion in the framework of the concept of
trapping or waiting times is given in Sec.\ref{single}. The stationary properties of
the many particle motion is prestented for non-zero average bias in Sec.\ref{many} and
for zero average bias in Sec.\ref{zero}. Some related models are discussed in Sec.\ref{related}
and non-stationary phenomena is described in Sec.\ref{non-stationary}. In the final section 
of this article we will summarize and discuss our results..

\section{Model}
\label{model}
We consider the partially asymmetric exclusion process (PASEP) with site-dependend 
hop rates on a one dimensional lattice. Each
lattice site $i$ can either be empty, i.e. $\tau_i=0$, or occupied by a single 
particle, $\tau_i=1$. 

The time evolution of the local configuration $(\tau_i, \tau_{i+1})$
is described by 
\beqn
(1, 0) &\to& (0, 1)
\quad \textrm{with rate } p_i,\\
( 0, 1) &\to& (1, 0)
\quad \textrm{with rate } q_i,
\eeqn
i.e. particles are hopping in positive direction (from 
$i$ to site $i+1$) with rate $p_i$ and in negative direction 
(from site $i+1$ to site $i$) with rate $q_i$. 

It is useful to relate the stochastic dynamics to a potential landscape 
by the relation
\be
\frac{q_i}{p_i} = e^{-(U_i-U_{i+1})},
\ee
where $U_i$ is the energy assigned to site $i$ (relative to a reference value, $U_1=0$). 
A typical sample of this type of landscape is shown in Fig. \ref{land}.  
Links with backward bias (${q_i}>{p_i}$) correspond to a descending 
links of the energy landscape while those with forward bias to ascending ones.
It turns out that the large scale
behavior of the process is determined by the wandering properties of the energy 
landscape. The relation between energy landscape and system properties
will be discussed in some detail in the next section.

The hop rates are independent and identically
distributed random variables taken from the distributions, $\rho
(p)dp$ and $\pi (q)dq$, respectively which will be specified later.
We restrict ourselves to types of randomness where there are both forward and backward links with finite probability, i.e. the easy
direction of hopping is also random.
We define a control parameter as 
the average asymmetry between forward and backward rates as follows:
\be
\delta=\frac{[\ln  p]_{\rm av}  - [\ln q]_{\rm  av}}{{\rm var}[\ln
p]+{\rm var}[\ln q]}\;,
\label{delta}
\ee
such that for $\delta>0$ ($\delta<0$) the particles move on average to
the right (left).  Here, and in the following $[\dots]_{\rm av}$
denotes average over quenched disorder, whereas ${\rm var}(x)$ stands
for the variance of $x$. Note, that $\delta \sim \lim_{L \to \infty} U_L/L$,
thus the control-parameter is proportional to the average slope
of the potential landscape.

In most of the calculations we use a binary disorder distribution:
\beqn
\rho (p)&=&c\delta (p-1)+(1-c)\delta(p-r) \nonumber \\
p_iq_i&=&r<1\quad \textrm{for all}~i\;,
\label{bimodal}
\eeqn
where the control-parameter takes the value:
\be
\delta=\frac{1-2c}{2c(1-c) \ln r}\;,
\label{delta1}
\ee
thus a biased motion to the right is realized for $1 \leq c < 1/2$.

In some cases we use a power-law distribution:
\beqn
\rho(p) = \frac{1}{D}\, p^{-1+1/D}&,& \qquad  \textrm{for } 0<p\leq 1; \label{eq:pow} \\ 
\pi(q)  = \frac{q_{0}^{-1/D}}{D}\, q^{-1+1/D}&,&\qquad  \textrm{for } 0<q\leq q_{0} \nonumber \\
\nonumber
\eeqn
where $D^2 =\textrm{var}[ \ln p]=\textrm{var}[ \ln q]$
measures the strength of disorder and $\delta=\ln q_0/(2 D^2)$.


\section{Single particle motion - the concept of trapping times}
\label{single}
We wish to introduce in the next section a simple phenomenological approach to the the biased exclusion process, which is based on the dynamics of the corresponding single-particle problem, known as Sinai walk\cite{sinai,im}. 
It is a thoroughly studied problem and here we recapitulate some of its properties which are necessary to understand to many-particle problem.  
For a periodic chain of length, $L$, the stationary
drift velocity, $v_0$, for a given realization of disorder is given  by\cite{derrida_v}:
\be
\frac{1}{v_0} = \frac{1}{L}\sum_{i=1}^{L} \frac{1}{p_i \Pi_{i+1}} \left(1-\prod_{i=1}^L \frac{q_i}{p_i} \right)^{-1}\;,
\label{v_0}
\ee
where
\be
\Pi_{i+1}=\left(1+\sum_{k=1}^{L-1}\prod_{j=1}^k\frac{q_{i+j-1}}{p_{i+j}}\right)^{-1}\;
\label{pers}
\ee
is the persistence probability\cite{ir} at site $i+1$. This quantity measures the
fraction of walks which start at site $i+1$ and pass the $i,i+1$ link first from
the left, i.e. after a complete tour along the chain\cite{redner}. 

\begin{figure}[h]
\includegraphics[width=0.9\linewidth]{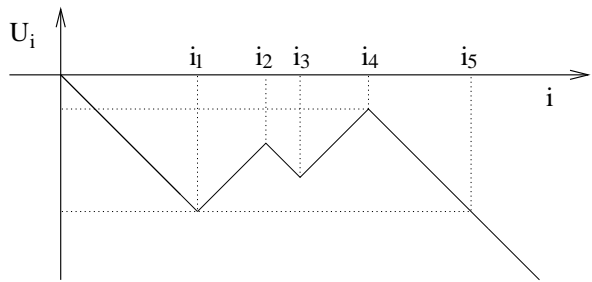}
\caption{\label{land0} Schematic energy landscape.}
\end{figure}

It is clear that the regions where the particle spends long times are
valleys in the energy landscape which are followed by a large barrier represented by a Brownian excursion, see the section $(i_1,i_5)$ in Fig. \ref{land0}.   
It is obvious that one can define a (single particle) trapping time for each barrier (Brownian excursion):
\be
\tau_{i}=\frac{1}{p_i \Pi_{i+1}}\;,
\label{trap}
\ee
which is the time needed to escape from that barrier and where $i$ is the starting point of the corresponding excursion, i.e. the minimal energy site of the valley in front of the barrier (site $i_1$ in Fig. \ref{land0}).   
The trapping time is of the form of a Kesten random variable\cite{kesten}, 
moreover one can convince oneself that trapping times belonging to different barriers are practically independent. It is known that in the large-$L$ limit the probability distribution of such variables  has an algebraic tail:
\be 
P(\tau ) \sim \tau^{-1-1/z},
\label{power}
\ee
where the exponent, $z$, is the positive root of the equation
\be
\left[\left({q\over p}\right)^{1/z} \right]_{av}=1.
\label{z}
\ee
Choosing the bimodal disorder as defined in Eq.(\ref{bimodal}) one obtains:
\be
z=\frac{\ln r}{\ln(c^{-1}-1)}\;
\label{z_bimodal}
\ee
Now concerning the drift velocity of the random walker it is 
the inverse of the average trapping time for $z<1$. If, however, $z>1$, the average of the
trapping time is divergent and the motion of a single particle
is determined by the largest trapping time, $\tau_{m}$, the typical value of
which follows from the relation\cite{galambos}: 
$\int_{\tau_m}^{\infty} P(\tau) {\rm d} \tau L=\mathcal{O}(1)$ and given
by $\tau_{m} \sim L^z$.

The persistence probability and thus the trapping time at a given barrier is related to the potential
landscape and for large trapping times it is asymptotically given by:
\be 
\tau_i \sim \sum_{k=1}^{L-1} \frac{1}{p_{i+k}}e^{U_{i+k}-U_{i}} \sim e^{\tilde{U}_i}
\label{tau}
\ee
which is well approximated by the largest term in the sum: $\tilde{U}_i= {\max_k}(U_{i+k}-U_{i})$.

In this way the motion of the particle for $z>1$ is
influenced only by the deep valleys and the corresponding large barriers, and
ultimately its velocity is determined by passing the largest barrier in the system, since
$\tau_m \sim e^{\tilde{U}_m}$ and $\tilde{U}_m= \max \tilde{U}_i$.
The intimate relation between the mobility of the particle and structure of the 
of the energy landscape led to the idea of renormalizing the energy landscape\cite{RGsinai,im}.
By using the renormalization 
scheme and alternative approaches one obtains the same form of the trapping time 
distribution\cite{cecile03} as given in Eq.~(\ref{power}).

\section{Many particle motion}
\label{many}

We now turn to the many particle case, i.e. we put $N=O(L)$
particles on the lattice.  First we consider a special form of disorder
for which exact results can be derived. These are then generalized for more general
form of disorder using phenomenological and scaling considerations.

\subsection{Extreme binary disorder}
\label{ext_disorder}

Here we consider the bimodal disorder as defined in Eq.(\ref{bimodal}) where the
landscape is represented by a random walk having a
step of size $\ln r$ with probability $(1-c)$ and of size $-\ln r$ with probability ($c$). 
In the limit $c \ll 1$
the walk is strongly biased, having typically downwards steps and the rare upward
steps are forming the barriers. 
If we also have $r \ll 1$ the large barriers are
typically straight, since for a large energy-scale the fluctuations are reduced\cite{RGsinai}.
The largest barrier
consists of $l_m$ upward steps, has a height of $\tilde{U}_m=l_m \ln r^{-1}$ 
and we have for the typical value $c^{l_m} L=1$. 
It is easy to check that the trapping time
for single particle motion is $\tau_{m} \sim r^{l_m} \sim L^{z'}$ and the dynamical
exponent $z'=\ln r/\ln c^{-1}$ corresponds to that in Eq.(\ref{z_bimodal}) in the given limit. 
\begin{figure}[h]
\includegraphics[width=0.9\linewidth]{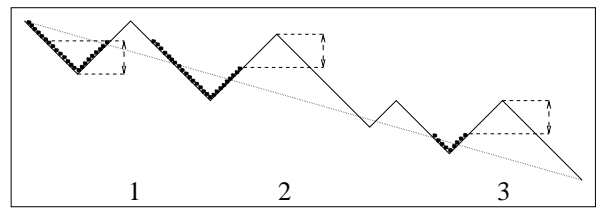}
\caption{\label{land}  Schematic form of the potential landscape for extreme disorder and
the typical position of the particles. The high-density and low-density phases are
separated at the middle of the largest barrier (2), the arrow indicates the height $U_m/2$.
Deep valleys in the low-density phase
are filled until the effective barrier is $U_m/2$ (3) and high barriers in the high-density
phase are vacant above a level of $U_m/2$ (1), see text.}
\end{figure}
Also for many particles the largest barrier plays the dominant role
and we analyze first the stationary current in a system having just one large barrier
of length $l_m$. 
This problem corresponds to a PASEP with particle in- and output 
against the direction of the bias. 
In the stationary state of this model, which is 
known exactly\cite{bece}, the system is half-filled and 
there is  a sharp front in the middle of the chain, which is the consequence of
particle-hole symmetry, so that particles (and holes) should overcome only an
effective (filled) barrier of height $\tilde{U}_m/2$. Consequently the effective trapping
time is $\tilde{\tau}_m \sim \tau_{m}^{1/2}$ and the stationary current is
given by:
\be
J \sim L^{-z/2}\;
\label{J}
\ee
which goes to zero in the thermodynamic limit.

Now, if we consider the complete potential landscape the basic features of the above
considerations remain true, since the largest barrier governs the stationary dynamics
of the system. In particular in the stationary state there is a sharp front located
at the middle of the largest barrier. Thus, due to particle-hole symmetry
the largest barrier is half-filled and the stationary current is related
to the largest effective trapping time and given by Eq.(\ref{J}).
The subleading barriers lead to a substructure both in the high-density and in the
low-density phases, which can be related to each other by using the particle-hole
symmetry. In the low-density region it is easy to see that small barriers with 
$\tilde{U}_i<\tilde{U}_m/2$ are completely empty, since they do not cause sufficient 
resistivity for the current. Only sufficiently large barriers with $\tilde{U}_i>\tilde{U}_m/2$
are able to slow down the particle and cause jams. These barriers will be filled up to the level:
$\tilde{U}_i-\tilde{U}_m/2$ as a consequence of the conservation of
the stationary current.
The typical number of occupied barriers in the low-density phase is given by: $c^{l_m/2} L \sim L^{1/2}$.
In the high-density phase due to particle-hole symmetry holes are accumulated only in such
subleading barriers for which $\tilde{U}_i>\tilde{U}_m/2$. The number of holes in such a barrier scales as
$\tilde{U}_i-\tilde{U}_m/2$ and the typical number of occupied barriers is $O(L^{1/2})$.

\subsection{General form of disorder}

Next we extend our discussion
to general form of disorder, when
it may be still assumed that there is a sharp front separating the
high-density and low-density regions.
If this is the case, the front is located at a position, where the escape rate of particles and the escape rates of holes which leave the front to the left through the high-density region, are equal, which is a necessary condition for the the localisation of the front.
As in the extreme disorder case, the front is located in the middle of the largest barrier..
The potential landscape compared with that for extreme disorder in Fig.\ref{land} is
modified in such a way that instead of long straight lines the barriers and valleys are
made by Brownian excursions.

\subsubsection{Distribution of the current}

The typical value of the stationary current follows from the same reasoning as in
Section \ref{ext_disorder}: it is given by
the square root of the trapping time of a single walker at the largest 
barrier. Furthermore the trapping time distribution follows  an asymptotically
algebraic distribution given by (\ref{power}).  

Therefore for a given system we have to find the maximum out of
$\mathcal{O}(L)$ algebraically distributed random variables,
a problem which is
thoroughly studied in the mathematical literature (see e.g.~in~[\onlinecite{galambos}])
and has demonstrated to work for strong Griffiths singularities in disordered systems\cite{jli06}.
For our case it follows that the current is described by the 
well-known Fr\'echet distribution given by
\be 
P(\tilde J)=\frac{2}{z}\tilde J^{2/z-1}e^{-\tilde J^{2/z}},
\label{frechet}
\ee
in terms of $\tilde J =J_0JL^{z/2}$, where the non-universal constant 
$J_0$ depends on the prefactor of the tail.  
Thus, the current scales with the system size as given in Eq.(\ref{J}).

Although the above results hold for generic hop rate distributions, a
caveat is in order concerning the type of randomness. 
Besides the bimodal one, 
one may also consider hop rate distributions where the support of the
the forward rates, $p$, does not have a positive lower bound, such as
for the power-law distribution in Eq.(\ref{eq:pow}). In this 
case it may happen that the current is limited by the lowest hop rate
$p_{min}\sim L^{-D}$, i.e. by a single slow link rather than by the maximal
amplitude of a Brownian excursion, which is given in Eq.(\ref{J}).
Clearly, the exponent which enters in (\ref{frechet}) and
as well as in the scaling relation of the current (\ref{J}) is now ${\rm max}\{z/2,D\}$.

\begin{figure}[h]
\includegraphics[width=0.8\linewidth]{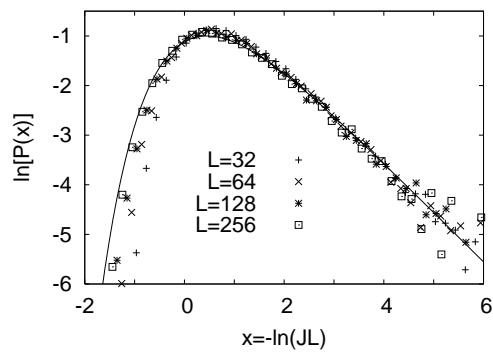}
\caption{\label{jdist} The distribution of the current, calculated
  with uniform randomness with $q_0=1/3$, where $z=1.335$. The number
  of samples is $10000$. The solid line is the Fr\'echet distribution
  given in (\ref{frechet}).}  
\end{figure}

For a uniform distribution with $D=1$ the current scales as $J(L)\sim L^{-z/2}$
for $z>2$ whereas $J(L)\sim L^{-1}$ for $1<z<2$. 
Numerical results for the distribution of current in the anomalous
region  $1<z<2$ are shown in Fig. \ref{jdist}, whereas numerical results for 
the bimodal randomness,
for which the anomalous scenario never sets in, can
be found in Ref\cite{zrp}.    
In what follows we assume that $z/2>D$ always holds and the 
current is controlled by an extended region. 

\subsubsection{Density profile}

The density profile for general form of disorder has the same qualitative features
as for the extreme disorder in Sec. \ref{ext_disorder}. The system consists of a high-density
phase, where all the lattice sites are typically occupied, and of a low-density
phase in which all the lattice sites are typically vacant. The front separating the
two phases is just at the middle of the largest barrier. In the low-density phase
only the large valleys with $U_l \geq U_m/2$ are filled to a level
$U_l - U_m/2$. Similarly, in the high-density phase only the large barriers with
$U_l \geq U_m/2$ are only partially filled up to $U_m/2$.

The number of occupied valleys $n(L)$, (and the number of partially filled barriers) can
be estimated from the condition, that at these barriers the single
particle trapping time is larger than $1/J$, thus
\be
n(L)\sim L\int_{1/J}^{\infty}p(\tau )d\tau \sim LJ^{1/z},
\label{n}
\ee
where we used (\ref{power}) and the fact that, due to their finite
typical extension, the number of barriers in the 
system is of order $L$. Keeping in mind the scaling relation of the current 
in (\ref{J}), we obtain $n(L)\sim L^{1/2}$, in agreement with the
extreme disorder case in Sec. \ref{ext_disorder}. This finding is in
good agreement with numerical results shown in Fig. \ref{cross}.

The length of Brownian excursions has an exponential
distribution\cite{ir97}, with a finite $\delta$-dependent
characteristic value. Although, the length of the $L^{1/2}$ longest excursions
is $\mathcal{O}(\ln L)$\cite{galambos}, the length of the empty domain in each of these barriers equal approximately to the half of the length of the largest barrier, so one can show that the number of particles 
they contain is typically finite.

\begin{figure}[h]
\includegraphics[width=0.8\linewidth]{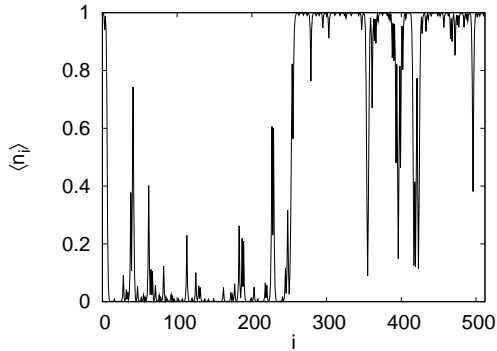}
\caption{\label{prof} Numerically calculated stationary density
  profile of a sample of size $L=512$. The hop rates were generated
  with the bimodal distribution with $c=0.3$ and $r=0.5$ in Eq.(\ref{bimodal}).}  
\end{figure}

\begin{figure}[h]
\includegraphics[width=0.8\linewidth]{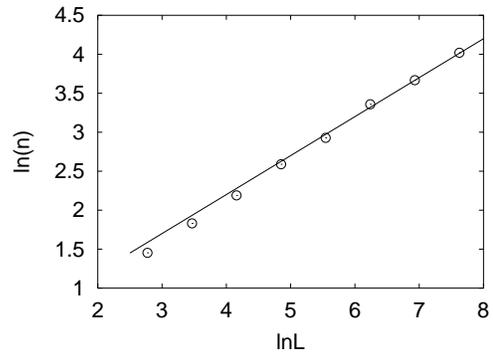}
\caption{\label{cross} Average number of points where the density
  profile crosses the line $\rho (x)=1/2$ calculated numerically for
  different system sizes $L$. Binary randomness were used with $r=0.5$
  and $c=0.2$, whereas the average was performed over 100 samples. The
  straight line has slope of $1/2$.}  
\end{figure}

The total number of particles clustering behind the subleading barriers
is thus of the order of $L^{1/2}$. The other $\mathcal{O}(L)$
particles must be behind the largest barrier, where a macroscopic
cluster of particles is formed.
According to the discussion given above, the particles can be
categorized into three classes as was carried out in a similar way for 
the disordered zero-range process\cite{zrp}.

Almost all particles are found  at the largest barrier,
forming a {\it condensate}, where the density is close to one.  
The length of the condensate and the number of particles in it 
is thus
\be
N_c \sim N.
\label{cond}
\ee
This region is, however, cut  into sections of size $\sim L^{1/2}$ by vacant clusters, which
have a finite length. 

In the remaining part of the system which is of size $\sim(L-N)$,
particle clusters of finite size are found, separated by a typical
distance of $L^{1/2}$. Due to the small local velocity in these
dense clusters, the
contribution of a single particle to the current is small compared to freely
moving particles, therefore we call these particles {\it inactive}. 
Their total number is given by 
\be 
N_{ia}\sim L^{1/2}.
\label{in}
\ee

The third class of particles, i.e. {\it active} ones,  
move practically freely between clusters of inactive particles.
Consequently, they have a significant contribution to the current.
In order to estimate the number of active particles, we notice that
due to the low particle density within sections between two
neighboring clusters of inactive particles the dynamics of active 
particles corresponds to the dynamics of a single random walker in 
a disordered landscape. The typical length of these section is
$\sim L^{1/2}$ and the largest trapping time of an 
active particle is $\sim \sqrt{\tau_1}\sim L^{z/2}$. Now if  $z<1$, 
the particle velocity $v$ is $L$-independent. Making use of the finite-size 
scaling form of the current, we obtain
for the density $\rho=J/v\sim L^{-z/2}$, which is indeed vanishing 
in the large $L$ limit. The total number of active particles is
thus 
\be
N_a\sim L^{1-z/2}, \qquad (z<1).
\label{a1}
\ee
For $z>1$, the largest barrier in the empty section determines the travel time 
 and the average velocity of particles scales according to
$v(L)\sim L^{(1-z)/2}$. The density  thus
scales as $\rho =J/v\sim L^{-1/2}$, which means that typically 
$\mathcal{O}(1)$ particles reside between two inactive clusters. In this 
case the  total
number of active particles is given by
\be 
N_a\sim L^{1/2}, \qquad (z>1).
\label{a2}
\ee
According to the discussion above, the distribution of interparticle 
distances in a finite
system of size $L$ has a finite cut-off: they have the scaling form
$\overline l\sim L^{z/2}$
for $z<1$, whereas $\overline l\sim L^{1/2}$  for  $z>1$.

Next, the number of particles at the second largest barrier $n_2$ is
analyzed. The height of the 2nd largest
barrier is proportional to $\ln L$ just as the height of the largest 
one.
For the bimodal randomness where the increments of the potential are
fixed, the length of the barrier must grow with the system size at least
as $\mathcal{O}(\ln L)$. On the other hand, it is generally true for
types of randomness under consideration,
where the current is controlled by an extended cluster of exponentially
rare links. 
The typical value for $\langle n_2\rangle$ is thus
$\mathcal{O}(\ln L)$. 
In certain samples, however, where the single-particle trapping
times  $\tau_1$ and  $\tau_2$ at the 1st and 2nd largest barriers,
respectively, are almost degenerate, meaning that $\tau_2\lesssim
\tau_1$, the occupation $\langle n_2\rangle$ is much larger, than the typical value $\mathcal{O}(\ln L)$.   
We show below, that these samples give a singular contribution to the
sample-averaged occupation $[\langle n_2\rangle]_{av}$.
In samples, where the two largest barriers are almost degenerate, 
$n_2$ has large-scale fluctuations and the dominant contribution to the thermal average $\langle n_2\rangle$
comes from the fluctuations where $n_2$ is much larger than its
most probable value. In the latter case half-filling is realized at the 2nd barrier, leading to
an escape time $\sqrt{\tau_2}$.
Not considering the improbable samples with
higher degeneracies (the contribution of which has in fact the same type
of singularity), the number of particles located at barriers other than the
two largest ones is negligible, and the problem is reduces to an
effective two-barrier process with hop rates $\sqrt{\tau_1}$ and $\sqrt{\tau_2}$.
Taking into account, that the total number of particles is $N$, we
obtain for the thermal average\cite{zrp}
\be
\langle n_2 \rangle (\alpha)={\alpha\over 
1-\alpha}-(N+1){ \alpha^{N+1}\over 1-\alpha^{N+1}},\quad \alpha<1\;,
\label{corr}
\ee
where $\alpha \equiv\sqrt{\tau_2/\tau_1}$, and $\langle n_2
\rangle=N/2$ for $\alpha=1$. 
Using the distribution function, $\rho(\alpha)$ we
can average over realizations:  
$[\langle n_2 \rangle ]_{\rm av}=\int_0^1 \langle n \rangle (\alpha)\rho(\alpha) {\rm d}\alpha$, which is dominated by the
contribution as $\alpha \to 1$, where $\rho(\alpha)$ has a finite
limiting value. Keeping in mind that the maximal value of $\langle n_2 \rangle$ is
$N/2$, we can write
\be
[\langle n_{2} \rangle ]_{\rm av} \approx
\int_0^{1-2/N} \frac{\alpha}{1-\alpha}\rho(\alpha) {\rm d}\alpha \sim \rho(1)
 \ln N.
\ee
Thus the rare samples, where the second largest
trapping time can be arbitrarily close to the largest one, lead to a
logarithmically diverging contribution to the sample-averaged
occupation at the 2nd barrier.  

\subsection{Low density limit}

Contrary to the previous section, where the global density $\rho =N/L$ was
finite, we study here the case, where the number of particles scales as $N\sim
CL^a$ for large $L$, with $0\leq a< 1$. This implies that the
particle density is vanishing algebraically according to 
$\rho(L) \sim CL^{a-1}$ as $L\to\infty$. 
In this case the number of active particles may be limited by $N$,
which influences the current. 

We have learned in the previous chapter, 
that the system has a limited capacity for storing
particles, given by $N_{a}+N_{ia}$, and the excess of particles is
driven to the macroscopic condensate. 
Evidently, as long as $N$ exceeds the capacity of the system,
$N_{a}+N_{ia}$, the excess particles accumulate in the condensate and
the largest barrier is haf filled. Consequently
the current remains the same as the one observed for constant density. 
This is, however, no longer true if $N$ is smaller then the capacity
of the system. 

For $z>1$, the capacity scales as $(N_{a}+N_{ia})\sim L^{1/2}$
according to (\ref{a2}) and (\ref{in}). Thus, for $a>1/2$ the current, the
number of active and inactive particles are not influenced, whereas 
$N_c\sim L^a$. 
However, if $a<1/2$, the condensate vanishes, and since one can show that 
$N_a$ is proportional to $N_{ia}$ for arbitrary current,
both must be proportional to  $L^a$. Therefore
according to (\ref{n}) the current is given by  
$J\sim (n(L)/L)^z \sim L^{-z(1-a)}$.

For $z<1$, the capacity is dominated by $N_{a}$, which scales
according to (\ref{a1}). Thus, as far as $a>1-z/2$, the number of
active and inactive particles and the current agrees with the case 
of a finite density, whereas the size of the
condensate scales as $N_c\sim L^a$. If $a<1-z/2$, there is no condensate present in the system
and the number of active particles is limited to $N_a\sim L^a$.
Therefore the corresponding current scales as $J\sim L^{-(1-a)}$. Then the number of
inactive particles scales as $N_{ia}\sim n(L)\sim L^{1-(1-a)/z}$ according to eq.~(\ref{n}), 
which is indeed a vanishing fraction of the active particles, which we have tacitly assumed.
According to the latter expression, if $z+a<1$, $N_{ia}\to 0$ as
$L\to\infty$, meaning that all particles are active.  

Setting $a=0$ in the scaling forms obtained
for the current we recover the exact results known for the Sinai walk.

\section{Zero average bias}
\label{zero}
If  the control parameter is zero $\delta =0$, the average tilt of
the potential landscape, 
and consequently the sample-averaged current is zero even for finite
systems. 
However, as a consequence of the fluctuations of randomness,  
the potential landscape of a single sample of size $L$ is
tilted, having a typical slope of
$L^{-1/2}$, in accordance to central limit theorem. 
Therefore the quantity of interest is the
magnitude of the current for a finite system. 

As the symmetric point is approached $\delta\to 0$, both the typical
length and amplitude of Brownian excursions are diverging as $\xi\sim
\delta^{-2}$ and $U_m\sim \delta^{-1}$, respectively. 
 According to central limit theorem
the typical amplitude is then of the order $L^{1/2}$, 
consequently the magnitude of the current vanishes rapidly 
with the system size as
\be
-\ln |J|\sim L^{1/2}.
\label{symm}
\ee 
Since the largest excursion covers an $\mathcal{O}(1)$ fraction of the
landscape, its internal structure has to be investigated in order to get some
insight into the properties of the density profile.     
We consider first a finite sample with $\delta_L\equiv \sum_{i=1}^L
(\ln p_i-\ln q_i)=0$, where the potential is single-valued and the
current is zero. In the stationary state of this sample the particles
practically occupy the $N$ lowest-potential sites, and 
the clusters of occupied sites are represented by Brownian
excursions. (These are not to be confused with
 barriers, which are excursions starting at extremal sites of the landscape.)     
The number of such excursions $n$ in a system of size $L$ is estimated as
follows. 
For $\delta =0$, the asymptotic distribution of their length is
\be
p_l\sim l^{-3/2},
\label{ipd}
\ee
thus the length of the largest one among $n$ excursions,
$l_{max}$, is
typically of the order $n^2$.
On the other hand, $n$ is related to $L$ via 
$n\int_1^{l_{max} }lp_ldl\sim L$, yielding 
$n\sim L^{1/2}$ and $l_{max}\sim L$. Thus, the potential landscape 
 contains  $\mathcal{O}(L^{1/2})$ excursions and the largest one of them
 is macroscopic. 
If  the landscape is tilted, but $L$ is large, one 
can still construct an approximate profile by cutting the
landscape by a straight line, the slope of which
corresponds to the tilt of the landscape. Then sites below the line  
are filled with particles.
Since the current as well as the typical tilt of
samples vanish as $L\to\infty$, the profile constructed in this way is
expected to be not much different from the true stationary profile.   

\begin{figure}[h]
\includegraphics[width=0.8\linewidth]{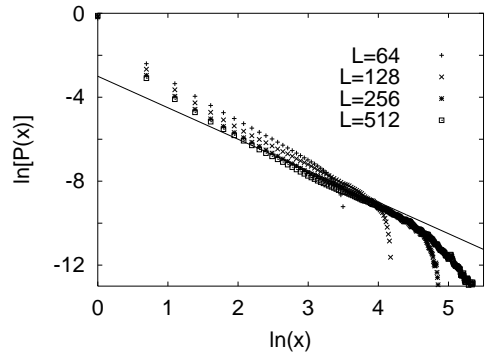}
\caption{\label{inter} Distribution of interparticle distances
  calculated numerically with the bimodal distribution with $c=0.5$
  and $r=0.2$. The number of samples is $5000$. The slope of the
  straight line is $-3/2$.}  
\end{figure}

According to the above one expects that the distribution of
interparticle distances has the asymptotics as given in (\ref{ipd}),
with a finite cut-off $\mathcal{O}(L)$. Indeed, numerical results are in 
satisfactory agreement with these predictions, see Fig. \ref{inter}.

\section{Related models}
\label{related}
In this section we shall apply the phenomenological description
developed above to two generalizations of the disordered exclusion process.   

\subsection{Particles of size $d$}

First, we consider a process, where the particles occupy 
$d\geq1$ subsequent lattice sites whereas they take steps
of unit length as before\cite{sethna}. 

\begin{figure}[h]
\includegraphics[width=1.1\linewidth]{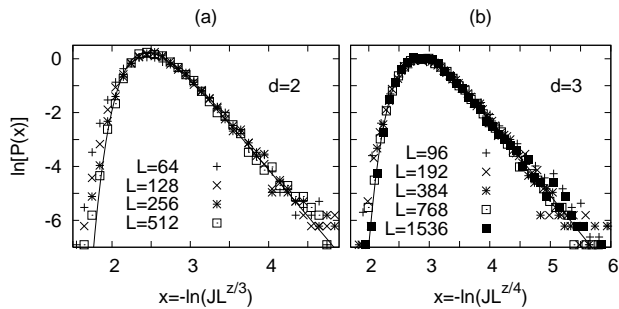}
\caption{\label{dfig} Distribution of the
  current for the exclusion process with particles of size $d=2$ (a)
  and $d=3$ (b), computed for different system sizes. The number of particles
  is $N=L/(2d)$.   
Bimodal randomness was used with $c=0.2$, $r=0.3$ where $z=0.869$ (a)
  and $c=0.3$, $r=0.3$, where $z=1.420$ (b), whereas the number of
  samples is $10^4$. The solid curves are the Fr\'echet
  distributions given in (\ref{frechet}).}  
\end{figure}

Obviously, the single-particle trapping time at a certain barrier,
 $\tau$ is independent of $d$. However,  
the time-scale for holes to overcome the same barrier 
which is now filled by particles of size $d$
is proportional to $\tau^{1/d}$, since the stepwidth of the holes 
is given by $d$, and the effective potential difference they feel is divided by $d$. 
Assuming that there is a sharp front at the largest barrier, it must be
 located at the potential where the escape rate for particles
 $\tau^{-1}$ and that for holes $\overline\tau^{-1}$ is related as 
$\overline\tau^{-1}=d\tau^{-1}$.
Parameterizing the escape rate as $\tau^{-1}=\tau_1^{-1/b}$, where
 $\tau_1$ is the single-particle trapping time, we have 
$\tau_1^{-(b-1)/(bd)}=d\tau_1^{-1/b}$. For large $\tau_1$ this yields 
$b=1+d$, consequently the current scales as 
\be 
J\sim L^{-z/(1+d)},
\label{Jd}
\ee
where we made use of the scaling relation $\tau_1\sim L^z$.
Following the argumentions of the previous section, 
similar results can be derived here, as well. 
For instance, the current follows a Fr\'echet distribution (see Fig
\ref{dfig} for
numerical results), and the number of inactive particles  scales 
with the system size $L$ as
\be
N_{ia}\sim L^{d/(1+d)}. 
\ee

Finally, we mention, that the model with $d=2$ is related to a PASEP in
which the particles are linked, meaning that the number of empty sites between  
two neighboring particles can be at most one.  
It is easy to see that the motion of holes in this model 
follows the same rules as, say, the left half of $d=2$ particles in the original model. 
In biological transport systems, molecular motors are often attached to a rigid backbone, 
where the distance between motors is limited to a finite value\cite{muscle}. Exclusion processes 
with linked particles may be relevant for modelling cooperative behavior in these type of systems. 

\subsection{Exclusion process with multiple occupation }

Our second example is the disordered version of the generalized  
exclusion process\cite{kipnis}, where the number of
particles at a given site is limited by the site independent integer 
$K\geq 1$. The top most particles at a given
site jumps to one of the neighboring sites with site-dependent rates
$p_i,q_i$, provided that the occupation at the target site is
smaller than $K$.  
This model evidently interpolates between the exclusion process
($K=1$) and the zero-range process ($K=N$). 
Moreover, it can be roughly interpreted as an exclusion process where the
size of particles, $d\equiv 1/K$, is smaller than one.

\begin{figure}[h]
\includegraphics[width=0.8\linewidth]{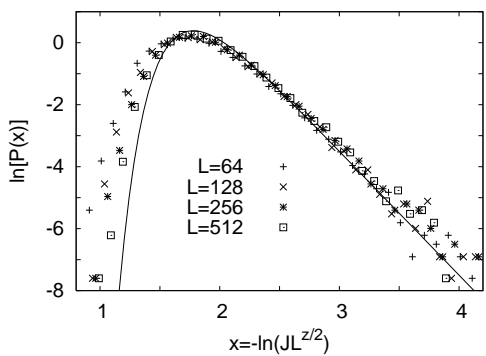}
\caption{\label{Kfig} Numerically calculated distributions of the
  current for the generalized exclusion process with $K=4$, for
  different system sizes. The number of particles is $N=KL/2$. Bimodal
  randomness was used with $c=0.2$ and $r=0.5$, where $z=0.5$, whereas
  the number of samples is $10^4$. The solid curve is the Fr\'echet
  distribution given in (\ref{frechet}).}  
\end{figure}

In the stationary state of this generalized model we still expect a 
front at the largest
barrier separating an almost fully occupied region (with $K$
particles per site) from an almost empty region. 
Moreover, particle-hole symmetry holds for $K>1$, as well,
therefore half-filling must be realized at the largest barrier, and
the current scales with the system size for any finite $K$ as for $K=1$.  
Numerical results for the distribution of the current are presented in
Fig. \ref{Kfig}.

One can also consider $L$-dependence in $K$, 
such as $K=K_0L^k$, with $0<k<1$, leading to a scaling of the
 number of inactive particles scales according to 
$N_{ia}\sim KL^{1/2}\sim L^{1/2+k}$. The results of the previous
section concerning the low-density limit can be generalized for this
case in a straightforward way.     

We mention finally, that regarding the number of particles
at a given site as a distance variable, the present model can be mapped to an
exclusion process with particlewise disorder, for any $K$, where the
distance between neighboring particles can be at most $K$.

\section{Non-stationary phenomena}
\label{non-stationary}
\subsection{Coarsening}

In the following we analyse the approach to 
the stationary state in an infinite system. 
When the system is started in a homogeneous configuration,
it undergoes a coarsening process in the course of which 
the typical size of high-density and low-density regions, $l(t)$, 
is growing \cite{krug}.
The growth rate of these regions is determined by the current
leading to the differential equation
\be 
{dl(t)\over dt} \sim |J(l(t))|. 
\label{diff}
\ee
where the current  $J(l(t))$ is a function of the time-dependent 
length scale.
By making use of the scaling relation of the current (\ref{J})
which is valid for $\delta \neq 0$ we get for large $t$ 
\be 
l(t)\sim t^{1/\zeta}, \quad \zeta ={z\over 2}+1, \qquad (\delta \neq
0) 
\label{zeta}
\ee
where $\zeta$ is the dynamical exponent related to the coarsening. 
We mention, that the same
scaling relation has already been derived for the bimodal
distribution, with $z'$ instead of $z$ \cite{krug}, which is therefore
valid only in the $c\to 0$ limit. Using (\ref{diff}) and (\ref{zeta}), 
the asymptotic time dependence of the current is given by  
\be
J(t)\sim t^{-{z\over 2+z}}, \qquad (\delta \neq 0).
\ee

\begin{figure}[h]
\includegraphics[width=0.9\linewidth]{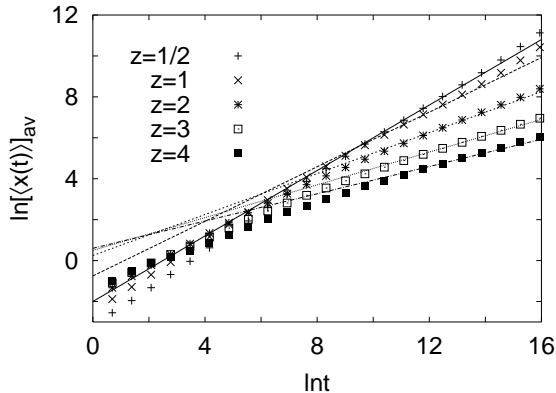}
\caption{\label{fig105} Time dependence of the average displacement of
  a single particle measured for different
  values of $\delta \neq 0$. Bimodal randomness was used with $c={1\over3}$ and
  $r={1\over \sqrt{2}},{1\over 2},{1\over 4}, {1\over 8},{1\over 16}$
  where $z={1\over 2},1,2,3,4$, respectively. The size of the system
  is $L=8192$ whereas $N=4096$ and the disorder average was performed
  over 200 samples. The straight lines have the slope ${2\over z+2}$
  for each $z$.  }  
\end{figure}

When the symmetric point is approached, i.e. $\delta\to 0$, the exponent
$z$ given in (\ref{z}), and consequently the dynamical exponent
$\zeta$ diverges as 
$\zeta \sim z \sim 1/(2\delta)$\cite{ir}. 
Thus, strictly at the symmetric point $\delta =0$, the length scale is
growing slower than any power of $t$, a phenomenon called anomalous
coarsening. 
Substituting the scaling relation of the current (\ref{symm}) into
(\ref{diff}), we get the asymptotic solution 
$l^{1/2}e^{{\rm C}l^{1/2}}\sim t$, leading to 
\be 
l(t) \sim 
\left[ \ln\left({t\over \ln t}\right) \right]^2, \quad  (\delta=0)
\label{log}
\ee   
which is indeed an anomalously (logarithmically) slow growth. 
Using (\ref{diff}) we obtain for the asymptotic time-dependence of the
magnitude of the current 
\be 
|J(t)|\sim 
{\ln t\over t}. \qquad (\delta =0) 
\ee
It is, however, circumstantial to measure the size of dense 
regions in numerical
simulations since they are not connected. Instead, we considered the 
average displacement of particles 
$\langle x(t)\rangle$ as a function of time for $\delta\neq 0$, which is related to the 
current according to $\frac{d\langle
  x(t)\rangle}{dt}=\frac{1}{\rho}J(t)$. Comparing this relation with
(\ref{diff}) we see, that $\langle x(t)\rangle$ grows in time as the
length scale $l(t)$.   
Results of numerical simulations shown in Fig. \ref{fig105}  are in good agreement with the phenomenological predictions.  

We mention, that the cut-off of the interparticle
distance distribution  $\overline l(t)$ is growing slower than $l(t)$. 
If  $z<1$, the interparticle distances in the low-density regions of
size $l(t)$ are proportional to $(l(t))^{z/2}$, and 
thus grow as $\overline l(t)\sim t^{z/(2+z)}$.
If $1<z<\infty$, the interparticle distances in the low-density regions are
proportional to $(l(t))^{1/2}$ leading to $\overline l(t)\sim t^{1/(2+z)}$.

\subsection{Invasion}

We consider in this section an open semi-infinite lattice with
entrance rate  $\alpha=1$, furthermore, it is assumed that 
the lattice is initially empty. Our interest is in the invading
dynamics, i.e. the dynamics of the first particle that enters 
the system and the motion of the bulk of particles. This question is of practical relevance,  e.g. one 
considers a fluid penetrating a porous medium. 
 
First, we study how the position of the leading particle 
evolves in time. 
Apparently, it must not travel slower than a single walker
since the jumps to the right are never hindered, contrary to jumps to
the left. 
Thus,  for $z<1$ the first particle advances with a constant velocity.  
For $z>1$, we consider a finite system of size $L$, and determine the
finite-size scaling of the characteristic time $\tau_p$ needed
for the first particle to reach site $L$, as follows.   
The leading contribution to $\tau_p$ comes from the trapping time at
the largest barrier and at subsequent barriers. Therefore we assume
that the first particle has already
 arrived at the largest barrier of the system. Without the other 
particles, the typical time to overcome the barrier 
would be $\tau_1\sim L^{z}$. However, during this period
other particles are arriving at the barrier with rate
$\mathcal{O}(L^{-z/2})$ determined by the current at the half-filled
largest barrier between the entrance site and leading particle.
The size of the largest barrier is at most $\sim \ln L$, so  
it takes at most a period of $t\sim L^{z/2}\ln L$ until half-filling of the largest barrier is achieved. 
Once half-filling is realized the trapping time for the leading particle is
only $\mathcal{O}(L^{z/2})$, so the total trapping time at the largest
barrier is at most $\mathcal{O}(L^{z/2}\ln L)$.

After leaving the largest barrier, the first particle advances freely until it
arrives at a barrier with trapping time larger then $\sqrt{\tau_1}$, where
it is assisted in passing the barrier again by other particles. 
The typical time-scale to fill the barrier is determined 
by the current and therefore given by $t\sim L^{z/2}$. Using the 
assistance of  the following particles the 
trapping time of the first particle at the barrier is 
given by $\mathcal{O}(L^{z/2})$. 
Since there are $\mathcal{O}(L^{1/2})$ barriers with trapping time 
$\tau > \sqrt{\tau_1}$ in a sample of size $L$, the characteristic time for
getting through the system scales in leading order as 
$\tau_p\sim L^{z/2+1/2}$. 
Therefore in a semi-infinite system, the position of the leading
particle $\xi_p$ is expected to scale as

\be
\xi_p \sim t^{1/z_p}, \quad {\rm with}  \quad z_p={z+1\over 2},
\label{inv}
\ee
for $z>1$, whereas $z_p=1$ for $z<1$.

\begin{figure}[h]
\includegraphics[width=1.1\linewidth]{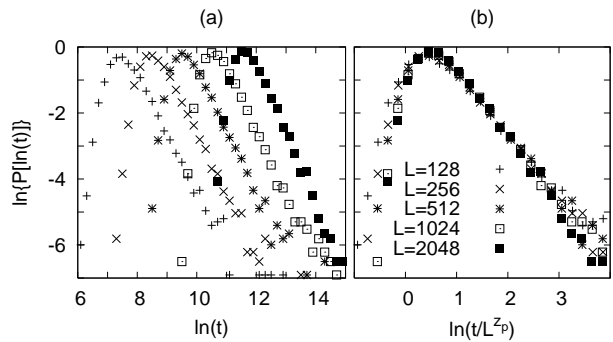}
\caption{\label{invt}  Distribution of invasion times calculated for
  different system sizes. Bimodal randomness was used with $c=0.3$ and
  $r=0.2$, where $z_p\approx 1.45$. The number of samples is $10^4$.}  
\end{figure}

In order to test the validity of our approach, we have checked the 
above results numerically by measuring the distribution of times the 
first particle needs to traverse a finite system. The results are
in good
agreement with the scaling relation (\ref{inv}), as shown in 
Fig. \ref{invt}. 
For zero average bias, the exponent $z_p$ is formally infinite and the
propagation of the first particle is anomalously slow.

\begin{figure}[h]
\includegraphics[width=0.9\linewidth]{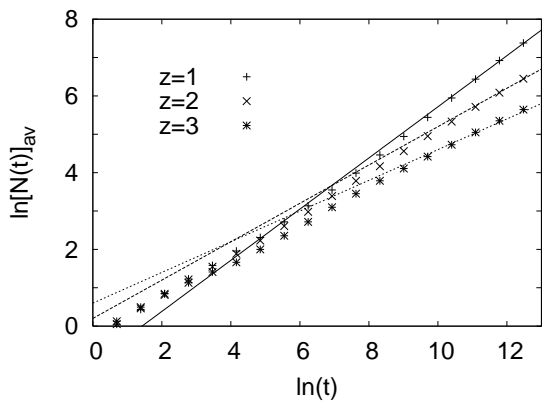}
\caption{\label{N} Time dependence of the average number of particles of
  measured for different
  values of $\delta \neq 0$. Bimodal randomness was used with $c={1\over3}$ and
  $r={1\over 2},{1\over 4}, {1\over 8}$
  where $z=1,2,3$, respectively. The size of the system
  is $L=8192$ and  the disorder average was performed
  over 200 samples. The straight lines have the slope ${2\over z+2}$
  for each $z$.}  
\end{figure}

Compared to the first particle, the bulk of particles moves slower, as we
shall show below. 
For large times, $t\gg 1$, a dense particle cluster is present in the system, 
which extends from the entrance site to a barrier which is located at $\xi_f(t)$ and which must be the largest barrier of the interval $[0,\xi_f(t)]$. 
Since this barrier is  half-filled, the current it maintains is typically 
in the order of $\xi_f(t)^{-z/2}$.
Through this (source) barrier
particles are constantly moving to the next larger barrier to its
right. The escape rate at this 2nd barrier is typically
$(2\xi_f(t))^{-z/2}$, which is a finite fraction of the filling
current, therefore the domain behind it is being filled up
by particles with a rate $\sim J\sim \xi_f(t)^{-z/2}$.
(Those particles which escape at the 2nd barrier meanwhile,
rush forward, to the third and subsequent larger barriers spreading 
in the domain behind the leading particle.) 
 After certain time the
domain between the source and the 2nd barrier is completely filled, and the new
position of the front is now at the 2nd barrier. 
Thus the growth rate of the dense cluster is proportional to the
current, $\frac{d\xi_f(t)}{dt}\sim J\sim \xi_f(t)^{-z/2}$, yielding
that the typical position of the front scales with the time as  
\be 
\xi_f(t)\sim t^{1/\zeta},        
\label{front}
\ee
with a larger dynamical exponent than that of the leading particle. 
In order to check
this numerically  we have measured the number of particles $N(t)$ as
 a function of time,
which is expected to grow as $\xi_f$. Results of numerical simulations
shown in Fig \ref{N}, are in good agreement with (\ref{front}).

\section{Discussion}

In this paper we have studied the partially asymmetric exclusion process with
lattice disorder and we have obtained several conjecturedly exact results both
for the stationary state and for related non-stationary problems. In particular for the
biased system we have shown a phase separation phenomenon where a domain wall  
taking place in the middle of the largest barrier separates a low- and a high-density phase. The stationary current in a large finite system goes
to zero as a power-law, and the corresponding exponent is just the half of the dynamical
exponent of a single random walker. We have also classified the particles, which typically
belong to a macroscopic condensate. This condensate, however, is cut up by $O(L^{1/2})$ vacant regions of
finite size, which are located at the subleading barriers of the potential landscape. These vacant
regions, through the particle-hole symmetry correspond to filled deep valleys in the
low-density phase, which contain the so called inactive particles. Finally, the third type of particles
are the active ones, which actually carry the current in the system. We have also studied the
properties of the system in non-stationary processes, such as in coarsening or in the invasion,
and have shown that the scaling exponents are related to the dynamical exponent of a single random walker.

Comparing the effect of different types of disorder on the stationary behavior of the PASEP we
can notice some analogies but also several differences. Both for particle and lattice disorder
the stationary current is vanishing as a power-law of the size of the lattice, and the critical
exponents are simple related to each other. Also the phase-separation phenomena is common for the
two problems. The system with particle disorder can be treated by renormalizing the
particles\cite{jsi} arriving to an effective model containing only a
single particle. Renormalization for lattice disorder has only been performed for a single
particle\cite{RGsinai} leading to a motion of the particle in a renormalized landscape. For many particles
it is expected that both the landscape and the particles should be renormalized and the resulting
model is still a many particle problem. The actual construction of the renormalization procedure
is the purpose of future research.

\section{Acknowledgments}

R.J. and L.S. acknowledge support by the Deutsche
Forschungsgemeinschaft under grant No. SA864/2-2. 
This work has been
supported by a German-Hungarian exchange program (DAAD-M\"OB), by the
Hungarian National Research Fund under grant No OTKA TO37323,
TO48721, K62588, MO45596 and M36803. F.I. is gratefull to C. Monthus for valuable discussions.



\end{document}